\documentclass[]{interact}

\usepackage{epstopdf}
\usepackage[caption=false]{subfig}

\usepackage[numbers,sort&compress]{natbib}
\usepackage{float} 
\usepackage{appendix}
\usepackage{import}
\usepackage{hyperref}
\usepackage{multirow}
\bibpunct[, ]{[}{]}{,}{n}{,}{,}

\theoremstyle{plain}

\theoremstyle{definition}

\theoremstyle{remark}

\begin{document}

\articletype{ARTICLE}

\title{A New Weighted Food CPI from Scanner Big Data in China}

\author{
\name{Zhenkun Zhou\textsuperscript{a,b}
, Zikun Song\textsuperscript{a,b}
and Tao Ren\textsuperscript{a,b,*}\thanks{*Corresponding auther: rentao@cueb.edu.cn} 
}
\affil{\textsuperscript{a}School of Statistics, Capital University of Economics and Business, Beijing, China\\
\textsuperscript{b}China Institute of Consumption Big Data, Capital University of Economics and Business, Beijing, China}
}

\maketitle

\begin{abstract}
Scanner big data has potential to construct Consumer Price Index (CPI). 
The study introduces a new weighted price index called S-FCPIw, which is constructed using scanner big data from retail sales in China. 
We address the limitations of China's CPI especially for its high cost and untimely release, and demonstrate the reliability of S-FCPIw by comparing it with existing price indices. S-FCPIw can not only reflect the changes of goods prices in higher frequency and richer dimension, and the analysis results show that S-FCPIw has a significant and strong relationship with CPI and Food CPI.
The findings suggest that scanner big data can supplement traditional CPI calculations in China and provide new insights into macroeconomic trends and inflation prediction.
We have made S-FCPIw publicly available\footnote{https://github.com/kayzhou/S-FCPIw} and update it on a weekly basis to facilitate further study in this field.
\end{abstract}


\section{Introduction}

Big data gives us an unprecedented opportunity to construct the real-time and high-frequency economic indicators, which provides a novel perspective for the economics research. The Consumer Price Index (CPI), being one of the key economic indices, measures inflation by monitoring price changes in a basket of goods and services commonly bought by households. It helps economists and policymakers understand the market while prices are rising or falling over time. However, China's CPI is derived and computed through limited surveys, and the figures are typically published on a monthly or quarterly basis.

Scanner data refers to structured information that includes product prices, goods categories, sales volume and store location. This data is generated when sellers scan the barcode of the goods and customers complete purchases at retail stores. It not only provides new data for constructing price indices but also enables indices at detailed aggregation levels~\citep{2011Eliminating}.
The Netherlands is pioneering the utilization of scanner data for CPI. Initially, researchers constructed the price index just for coffee~\citep{1997Estimation}. In Norway and New Zealand, scanner data is officially used for CPI~\citep{nygaard2010chain,ivancic2011scanner}.
Due to concerns regarding data security and privacy, retail enterprises lack of motivation to provide scanner data. Consequently, collecting and obtaining scanner data becomes very challenging. Some researchers have turned to study available online price data. Notably, there are two well-established research projects: the Billion Price Project conducted by MIT~\citep{cavallo2016billion}, and the iCPI developed by researchers from Tsinghua University for China~\citep{LiuTaoxiong2019}.

To our best knowledge, there is currently no research that directly employs scanner data to construct CPI or Food CPI in China. Consequently, we undertake an analysis of scanner data from the China Ant Business Alliance, especially for food prices. We establish an weighted index S-FCPIw and compare it with existing price indices. This work paves a new way for real-time monitoring change of price, reflecting macroeconomic trends and measuring inflation inertia, and is a valuable supplement to the research of CPI.

\section{Data}

The data utilized to construct price index originates from the retail scanner database provided by China Ant Alliance (CAA), which is a retail alliance organization in China. The CAA has more than 100 member enterprises in 29 provinces in China, with a total annual turnover approaching 100 billion Yuan\footnote{http://en.mayishanglian.com/}. By utilizing the Alibaba Cloud platform, the CAA has been collecting and processing the massive scanner data from the member enterprises in real time.

Compared with survey, scanner data in our study offers three main advantages: low cost, high frequency and rich dimensions. Here, we further elaborate on the rich dimensions. The scanner database encompasses the sales data of 12 million products spanning 29 provinces in China. This allows us to construct an index at city, province and nation level in terms of regional dimension. Regarding the dimension of goods categories, it include 75 basic categories and 14 sub-categories.

\section{Method}

We delineate the process of constructing S-FCPI and S-FCPIw below. Given that both S-FCPI and S-FCPIw share the same calculations in the majority of steps, we use the S-FCPI to denote both indices.

\textbf{Classifying goods.} 
The official criteria for goods classification, such as the COICOP and the Statistical Report on Circulation and Consumer Prices in China, are employed to reclassify the categories of goods in the scanner database (Table~\ref{tab:categories}).

\textbf{Obtaining goods price.}
The selection of representative goods for each category is facilitated by the local government based on sales volume and consumption structure, thereby creating a relatively fixed basket of goods. In contrast, S-FCPI incorporates almost all goods from the scanner data, utilizing the product ID to correlate sales data across different periods. Consequently, we create a variable basket of goods.
Regarding the frequency of price collection, the survey for CPI is gathered between one to five times per month. However, each recorded purchase can be regarded as price collection of the goods listed on the receipt. The availability of scanner data enables real-time price collection, allowing the S-FCPI to be constructed not only monthly, but also at a higher frequency.

\textbf{Calculating weight.}
Since the categories of goods account for different proportions, it is necessary to set corresponding weight for categories. The weight in CPI is derived throughout the survey, and is slightly adjusted in each year. In addition, CPI has no weight below the basic-category. The weight in S-FCPI is calculated using real-time scanner data which enables timely adjustment during the report period. Specifically, we select the corresponding aggregate scanner data based on the time and region dimension, then calculate the proportions of goods sales to determine the weights.

\textbf{Calculating index.}
\textbf{Step one, calculating the unit price.} 
The scanner data provides comprehensive information including both price and sales volume of the stock keeping unit (SKU). By using the weighted average SKU's price (unit price) instead of the simple average used by CPI, we can accurately reflect the differences between stores.
We assign the share of sales volume of goods as the weight and calculate the weighted average price across multiple stores.
$U_k^t$ represents the price of SKU $k$, $Sale_{k,e}^t$ represents total sale volume at the store $e$, and $Q_{k,e}^t$ represents total sale amount at the store $e$ and time slot $t$.
\begin{equation}
U_k^t=\frac{\sum Sale_{k,e}^t}{\sum Q_{k,e}^t}\label{eq:U}.
\end{equation}
\textbf{Step two, calculating price changes.} 
The relative value $R_k^t$ of price changes of each SKU in S-FCPI is calculated by comparing the unit prices of the two slot $t$ and $t-1$.
\begin{equation}
R_k^t=\frac{U_k^t}{U_k^{t-1}}\label{eq:R}.
\end{equation}
\textbf{Step three, constructing the price index for the basic-category.} In China, the National Statistic Bureau adopts the fixed basket index theory~\cite{XU2006cpi} and uses the Jevons price index, as China's official CPI. It does not take into account consumers' preferences. Referring to CPI, we create the Scanner-based Food CPI (S-FCPI). Regrading the basic-category $j$ and the sub-category $i$, S-FCPI is calculated by
\begin{equation}
\text{S-FCPI}_j^t
=\prod_{k=1}^{N_{jk}^t}\left(R_{jk}^t\right)^\frac{1}{N_{jk}^t}\label{eq:J},
\end{equation}
where $N_{jk}^t$ is the number of goods in the category $j$.
The above form is often used when the weights are unavailable. The scanner data makes it possible to construct the weighted price index for basic-category, considering consumer preference.
The Consumer Price Index Manual also points out that the weighted index will be a preferable option for  CPI~\cite{graf2020consumer}.
Therefore, inspired by the theory of aggregator function in T\"ornqvist index~\cite{diewert1976exact}, we construct a new weighted food index in China called S-FCPIw, 
\begin{equation}
\text{S-FCPIw}_j^t
=\prod_{k=1}^{N_{jk}^t}\left(R_{jk}^t\right)^\frac{S_{jk}^{t-1}+S_{jk}^t}{2}\label{eq:T},
\end{equation}
\begin{equation}
S_{jk}^t
=\frac{{Sales}_{jk}^t}{\sum_{k=1}^{N_{jk}^t}{Sales}_{jk}^t},
\end{equation}
where $S_{jk}^t$ is the share of sales of $k$ in the basic-category $j$.
\textbf{Step four, aggregate to the high-level index.} China's CPI uses the Laspeyres price index for high-level aggregate, and its weights are based on the household consumption from the low-frequency survey. Therefore, S-FCPI takes the proportion of sales in the reporting period as the weight and employ the Paasche price index to aggregate the index~\cite{diewert1998index}.
First, S-FCPI for the basic-category at the city-level is aggregated at the sub-category city-level S-FCPI, and then aggregated into the all-category city-level S-FCPI.
Second, we aggregate the index at basic-category province-level, sub-category province-level and all-category province-level.
Third, the S-FCPI is aggregated at the nation-level for the basic-category, sub-category, and all-category. 
Thus, S-FCPI and S-FCPIw are obtained across three region dimension and three category dimension.

With the help of scanner big data, we can follow the method to obtain daily, weekly and monthly S-FCPI and S-FCPIw. We further investigate the nation-level all-category S-FCPI/S-FCPIw in the following.

\section{Results}

S-FCPI/S-FCPIw have significant advantages of low cost, high frequency and rich dimension. Both indices can help to reflect macroeconomic trends, measure inflation inertia and predict inflation. We examine the reliability of the new indices by investigating their relationship with existing indices. The CPI, FCPI (Food CPI) and FRPI (Food Retail Price Index) published by the National Bureau of Statistics of China are selected to analyze and compare in our study.
We compare S-FCPI/S-FCPIw with the online food price index released by the iCPI project.

\subsection{Co-directional rate analysis}

\begin{figure*}[]
\centering
 \includegraphics[scale=.45]{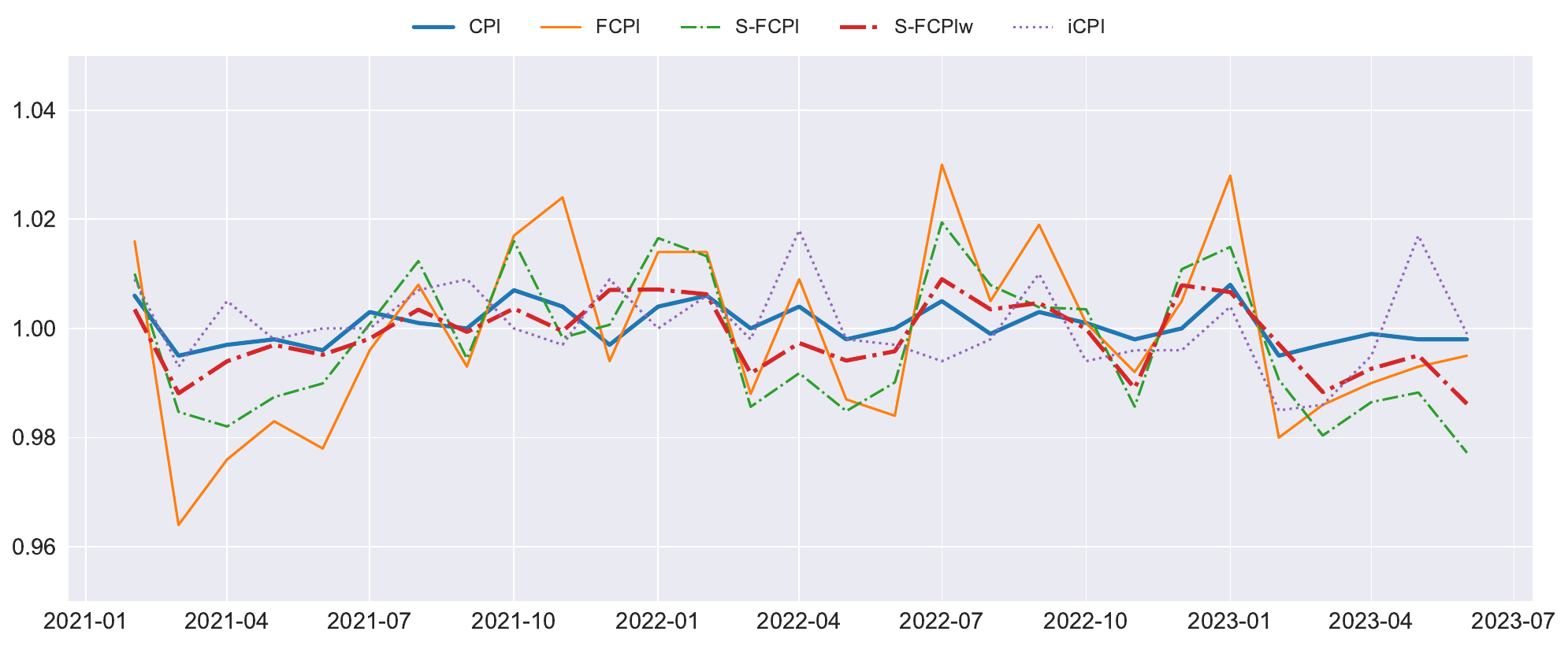}
 \caption{Month-on-month indices from 2021 to 2023.}
 \label{fig:line}
\end{figure*}

The index can be used to measure the change of price during the reporting period compared to the previous period. To evaluate the ability of S-FCPIw to reflect price fluctuations, we present the month-on-month indices in Fig.~\ref{fig:line} from January 2021 to June 2023.

To quantitatively measure the change direction of indices, we calculate the co-directional rates and present the results in Table~\ref{tab:corate}. The co-directional rate is calculated by $=\frac{m}{M}\times100\%$, where $m$ is the number of months with the same change direction between indices, and $M$ is the total number of months.

\begin{table}[H]
\centering
\caption{Co-directional rate (\%) between indices.}
\begin{tabular}{lcclc}
\toprule
 & CPI & FCPI &  FRPI & iCPI\\
\midrule
S-FCPI & 68.97 & 86.21 &  69.57 & 72.22\\
S-FCPIw & 75.86 & 86.21 &  69.57 & 66.67\\
iCPI & 65.52 & 72.22 &  60.87 & - \\
\bottomrule
\end{tabular}
\label{tab:corate}
\end{table}

The change direction between S-FCPI/S-FCPIw and FCPI are quietly close, with a co-directional rate of 86.21\%. They have the same change direction in 25 months and different directions only in 4 months. The co-directional rate of iCPI and CPI is lower than that of S-FCPI/S-FCPIw. Besides, S-FCPIw and CPI changed with the same direction in 22 months, accounting for 75.86\%. The co-directional rate between iCPI and CPI is the lowest (65.52\%). The results regarding FRPI are consistent with the above.

\subsection{Correlation analysis}

We further study the correlation between S-FCPIw and other indices.
The index growth rate is usually used to analyze the correlation between two chain indices. Therefore, we compute the logarithmic growth rate of each indicator's chain index through ${x_{rate, t}}=\ln{x_t}-\ln{x_{t-1}}$ and then analyze the Spearman correlation (Table~\ref{tab:corr}).

\begin{table}[H]
\centering
\caption{Spearman correlation between indices. Note: $^{***}$, $^{**}$ and $^{*}$ are significant at the level of 1\%, 5\% and 10\% respectively. The two rows for each pair of indices represent correlation coefficient and p-value.}
\begin{tabular}{lcccc}
\toprule
                         & CPI               & FCPI              & FRPI         &iCPI\\ \midrule
\multirow{2}{*}{S-FCPI}  & 0.5009$^{*}$    & 0.5524$^{*}$ & 0.5804$^{**}$    &0.5664$^{*}$\\
                         & 0.0972            & 0.0625         & 0.0479         &0.0548\\ \midrule
\multirow{2}{*}{S-FCPIw} & 0.6340$^{**}$    & 0.7063$^{**}$ & 0.7343$^{***}$   &0.2867\\
                         & 0.0268 & 0.0102 & 0.0065        &0.3663\\ \midrule         
\multirow{2}{*}{iCPI}    & 0.2837            & 0.3776            & 0.3776    &-\\
                         & 0.3715            &  0.2262            & 0.2262   &-\\ \bottomrule
\end{tabular}
\label{tab:corr}
\end{table}

S-FCPI demonstrates a strong positive correlation with CPI (coef=0.5009$^{*}$), FCPI (0.5524$^{*}$), and FRPI (0.5804$^{*}$), indicating significant relationship between S-FCPI and these indicators. 
Surprisingly, we find that the correlation for S-FCPIw is more significant than that of S-FCPI. S-FCPIw exhibits a strong positive correlation with CPI (0.6340$^{**}$), FCPI (0.7063$^{**}$) and FRPI (0.7343$^{**}$). 
However, we do not find a significant relationship between iCPI and other CPIs, including CPI, FCPI and FRPI. This implies that iCPI could not reflect the change in CPI while considering the correlation.

\section{Conclusion}

In this study, we demonstrate the construction method of a new weighted price index using scanner big data. Our work addresses the deficiency of China's CPI. Through co-directional rate and correlation analysis, we find that S-FCPIw has a significant relationship with price indices in China. However, S-FCPI has certain inherent limitations. Researchers can conduct further research in these promising directions. Regarding the theory, we could explore the issues of chain drift and transcendental index through scanner big data. 
In terms of index construction, S-FCPI can expand categories of goods and consider online price data. Based on the new index, researchers could further study the measurement of inflation inertia, analyze the stickiness of goods prices, and examine the heterogeneity of goods price fluctuating among regions.

\section*{Disclosure statement}

The authors report there are no competing interests to declare.

\section*{Funding}

This work was supported by National Social Science Fund of China (Grant No. 19BGL076), National Natural Science Foundation of China (Grant No. 72002141), Beijing Social Science Fund (Grant No. 20GLA074), R\&D Program of Beijing Municipal Education Commission (Grant No. KM202210038002), Special Fund for Fundamental Scientific Research of the Beijing Colleges in CUEB (Grant No. QNTD202109).


\bibliographystyle{model1-num-names}
%

\begin{thebibliography}{10}
\expandafter\ifx\csname natexlab\endcsname\relax\def\natexlab#1{#1}\fi
\providecommand{\url}[1]{\texttt{#1}}
\providecommand{\path}[1]{#1}
\providecommand{\DOIprefix}{doi:}
\providecommand{\ArXivprefix}{arXiv:}
\providecommand{\URLprefix}{URL: }
\providecommand{\Pubmedprefix}{pmid:}
\providecommand{\doi}[1]{\href{http://dx.doi.org/#1}{\path{#1}}}
\providecommand{\Pubmed}[1]{\href{pmid:#1}{\path{#1}}}
\providecommand{\bibinfo}[2]{#2}
\ifx\xfnm\relax \def\xfnm[#1]{\unskip,\space#1}\fi
\bibitem[{Haan and Van(2011)}]{2011Eliminating}
\bibinfo{author}{J.~D. Haan}, \bibinfo{author}{H.~A. Van, der~Grient},
\newblock \bibinfo{title}{Eliminating chain drift in price indexes based on scanner data},
\newblock \bibinfo{journal}{Journal of Econometrics} \bibinfo{volume}{161} (\bibinfo{year}{2011}) \bibinfo{pages}{36--46}.
\bibitem[{Haan and Opperdoes(1997)}]{1997Estimation}
\bibinfo{author}{J.~D. Haan}, \bibinfo{author}{E.~Opperdoes},
\newblock \bibinfo{title}{Estimation of the coffee price index using scanner data: Simulation of official practices},
\newblock \bibinfo{journal}{Simulation of Official Practices", Third Meeting of the International Working Group on Price Indices, Stats}  (\bibinfo{year}{1997}) \bibinfo{pages}{16--18}.
\bibitem[{Nygaard(2010)}]{nygaard2010chain}
\bibinfo{author}{R.~Nygaard},
\newblock \bibinfo{title}{Chain drift in a monthly chained superlative price index},
\newblock \bibinfo{journal}{article presented at the joint UNECE/ILOs workshop on scanner data, Geneva, Switzerland}  (\bibinfo{year}{2010}).
\bibitem[{Ivancic et~al.(2011)Ivancic, Diewert, and Fox}]{ivancic2011scanner}
\bibinfo{author}{L.~Ivancic}, \bibinfo{author}{W.~E. Diewert}, \bibinfo{author}{K.~J. Fox},
\newblock \bibinfo{title}{Scanner data, time aggregation and the construction of price indexes},
\newblock \bibinfo{journal}{Journal of Econometrics} \bibinfo{volume}{161} (\bibinfo{year}{2011}) \bibinfo{pages}{24--35}.
\bibitem[{Cavallo and Rigobon(2016)}]{cavallo2016billion}
\bibinfo{author}{A.~Cavallo}, \bibinfo{author}{R.~Rigobon},
\newblock \bibinfo{title}{The billion prices project: Using online prices for measurement and research},
\newblock \bibinfo{journal}{Journal of Economic Perspectives} \bibinfo{volume}{30} (\bibinfo{year}{2016}) \bibinfo{pages}{151--78}.
\bibitem[{Taoxiong et~al.(2019)Taoxiong, Ke, Tingfeng, and Li}]{LiuTaoxiong2019}
\bibinfo{author}{L.~Taoxiong}, \bibinfo{author}{T.~Ke}, \bibinfo{author}{J.~Tingfeng}, \bibinfo{author}{Z.~Li},
\newblock \bibinfo{title}{Design and application of novel cpi based on online big data},
\newblock \bibinfo{journal}{Journal of Quantitative \& Technological Economics} \bibinfo{volume}{36} (\bibinfo{year}{2019}) \bibinfo{pages}{21}.
\bibitem[{Qiang(2006)}]{XU2006cpi}
\bibinfo{author}{X.~Qiang},
\newblock \bibinfo{title}{The theoretical framework of cpi: Fixed-fixed-basket index or cost-of-living index},
\newblock \bibinfo{journal}{Research on Financial and Economic Issues}  (\bibinfo{year}{2006}) \bibinfo{pages}{19--27}.
\bibitem[{Graf(2020)}]{graf2020consumer}
\bibinfo{author}{B.~Graf},
\newblock \bibinfo{title}{Consumer price index manual, 2020: Concepts and methods},
\newblock in: \bibinfo{booktitle}{Consumer Price Index Manual, 2020}, \bibinfo{publisher}{International Monetary Fund}, \bibinfo{year}{2020}.
\bibitem[{Diewert(1976)}]{diewert1976exact}
\bibinfo{author}{E.~Diewert},
\newblock \bibinfo{title}{Exact and superlative index numbers, journal of econometrics, vol. 4}  (\bibinfo{year}{1976}).
\bibitem[{Diewert(1998)}]{diewert1998index}
\bibinfo{author}{W.~E. Diewert},
\newblock \bibinfo{title}{Index number issues in the consumer price index},
\newblock \bibinfo{journal}{Journal of Economic Perspectives} \bibinfo{volume}{12} (\bibinfo{year}{1998}) \bibinfo{pages}{47--58}.

\end{thebibliography}

\clearpage
\appendix

\section{Supplementary material}

\begin{table*}[h]
\centering
\small
\caption{Food basic-category and sub-category for S-FCPI.}
\label{tab:categories}
\begin{tabular*}{0.9\textwidth}{ll}
\toprule
sub-category                  &  basic-category                       \\ \midrule
Grain                         & Rice, noodles, other grains, and grain products                                                                                                                                                                 \\ \midrule
Tuber                         & Tuber, Tuber products                                                                                                                                                                                           \\ \midrule
Bean                          & Dried beans, soy products                                                                                                                                                                                       \\ \midrule
Edible oil                    & \begin{tabular}[c]{@{}l@{}}Rapeseed oil, soybean oil, peanut oil, sunflower oil,\\ camellia oil, blend oil, linseed oil, corn oil,\\ olive oil, butter, other edible oils\end{tabular}                           \\ \midrule
Vegetables and edible fungi   & \begin{tabular}[c]{@{}l@{}}Onion, ginger, garlic and pepper, root vegetables, \\ mushroom vegetables, nightshade vegetables, leafy \\ vegetables, dried vegetables and dried bacteria \\ and products\end{tabular} \\ \midrule
Neat of animal                & \begin{tabular}[c]{@{}l@{}}Pork, beef, mutton, other livestock meat and \\ by-products, livestock meat products\end{tabular}                                                                                    \\ \midrule
Meat of poultries             & Chicken, duck, other poultry meat and products                                                                                                                                                                  \\ \midrule
Aquatic products              & \begin{tabular}[c]{@{}l@{}}Fish, shrimp, crab, shellfish, algae, soft pod, \\ other aquatic products\end{tabular}                                                                                               \\ \midrule
Eggs                          & \begin{tabular}[c]{@{}l@{}}Eggs, duck eggs, goose eggs, quail eggs, pigeon eggs, \\ other egg products and products\end{tabular}                                                                                \\ \midrule
Dairy                         & \begin{tabular}[c]{@{}l@{}}Pure milk, pure goat milk, yogurt, milk powder, \\ milk beverage, other milk products\end{tabular}                                                                                   \\ \midrule
Dried fresh melons and fruits & \begin{tabular}[c]{@{}l@{}}Fresh fruit, nuts, candied dried fruit, \\other melon and fruit products\end{tabular}                                                                                                                                          \\ \midrule
Candy and pastries            & \begin{tabular}[c]{@{}l@{}}Sugar, sweets and chocolate, pastries, \\other confectionery pastries\end{tabular}                                                                                                                                             \\ \midrule
Condiments                    & \begin{tabular}[c]{@{}l@{}}Edible salt, soy sauce, vinegar, cooking wine, \\chicken essence, monosodium glutamate, sesame oil, \\ seasoning sauce, chili oil,  spices, other condiments\end{tabular}            \\ \midrule
Other food categories         & \begin{tabular}[c]{@{}l@{}}Convenience food, starch, and products, \\puffed food, baby food\end{tabular}                                                                                                                                                  \\ \bottomrule
\end{tabular*}
\end{table*}

\end{document}